\documentclass[twocolumn]{aastex62}

\newcommand{\eg}{e.g.,\ }

\newcommand{\Msun}{$M_{\odot}$}

\newcommand{\kms}{km~s$^{-1}$}

\newcommand{\TiII}{Ti~{\sc ii}}

\newcommand{\FeII}{Fe~{\sc ii}}

\newcommand{\CoII}{Co~{\sc ii}}

\newcommand{\NiII}{Ni~{\sc ii}}

\newcommand{\Nifs}{$^{56}$Ni}

\newcommand{\sBV}{s$_{BV}$}

\newcommand{\ab}{$\sim$}
\newcommand{\mic}{$\mu$m}
\newcommand{\DmB}{\Delta{ \rm{m}_{15}(B)}}

\newcommand{\ved}{$v_{edge}$}

\graphicspath{{./}{figures/}}

\received{\today}
\revised{}
\accepted{}
\submitjournal{ApJL}

\shorttitle{\ved\ probing \Nifs}
\shortauthors{Ashall et al.}

\begin{document}

\title{\textit{Carnegie Supernova Project-II: } 
Using Near-Infrared Spectroscopy to determine the location of the outer \Nifs\ in Type Ia Supernovae\footnote{This paper includes data gathered with the 6.5 meter Magellan Telescopes located at Las Campanas Observatory, Chile}}

\correspondingauthor{Chris Ashall}
\email{Chris.Ashall@gmail.com}

\author{C. Ashall}
\affil{Department of Physics, Florida State University, Tallahassee, FL 32306, USA}

\author{E. Y. Hsiao}
\affiliation{Department of Physics, Florida State University, Tallahassee, FL 32306, USA}

\author{P .Hoeflich}
\affiliation{Department of Physics, Florida State University, Tallahassee, FL 32306, USA}

\author{M. Stritzinger }
\affiliation{Department of Physics and Astronomy, Aarhus University, 
Ny Munkegade 120, DK-8000 Aarhus C, Denmark }

\author{M. M. Phillips}
\affiliation{Department of Physics, Florida State University, Tallahassee, FL 32306, USA}
\affiliation{Carnegie Observatories, Las Campanas Observatory, 601 Casilla, La Serena, Chile }

\author{N. Morrell}
\affiliation{Carnegie Observatories, Las Campanas Observatory, 601 Casilla, La Serena, Chile 
}

\author{S. Davis}
\affiliation{Department of Physics, Florida State University, Tallahassee, FL 32306, USA}

\author{E. Baron}
\affiliation{Department of Physics and Astronomy, Aarhus University, 
Ny Munkegade 120, DK-8000 Aarhus C, Denmark }
\affiliation{Homer L. Dodge Department of Physics and Astronomy, University of Oklahoma, 440 W. Brooks, Rm 100, Norman, OK 73019-2061, USA}
\affiliation{Hamburger Sternwarte, Gojenbergsweg 112, D-21029 Hamburg, Germany}

\author{A. L. Piro}
\affiliation{Observatories of the Carnegie Institution for Science, 813 Santa Barbara St., Pasadena, CA 91101, USA }

\author{C. Burns}
\affiliation{Observatories of the Carnegie Institution for Science, 813 Santa Barbara St., Pasadena, CA 91101, USA }

\author{C. Contreras }
\affiliation{Carnegie Observatories, Las Campanas Observatory, 601 Casilla, La Serena, Chile}

\author{L. Galbany}
\affiliation{PITT PACC, Department of Physics and Astronomy, University of Pittsburgh, Pittsburgh, PA 15260, USA}

\author{S. Holmbo}
\affiliation{Department of Physics and Astronomy, Aarhus University, 
Ny Munkegade 120, DK-8000 Aarhus C, Denmark }

\author{R. P. Kirshner}
\affiliation{Gordon and Betty Moore Foundation, 1661 Page Mill Road, Palo Alto, CA 94304}
\affiliation{Harvard-Smithsonian Center for Astrophysics, 60 Garden Street, Cambridge, MA 02138}

\author{K. Krisciunas }
\affiliation{George P. and Cynthia Woods Mitchell Institute for Fundamental Physics \& Astronomy, Texas A\&M University, Department of Physics, 4242 TAMU, College Station, TX 77843}

\author{G.~H.~Marion}
\affiliation{University of Texas at Austin, 1 University Station C1400, Austin, TX, 78712-0259, USA}

\author{D. J. Sand}
\affil{Department of Astronomy/Steward Observatory, 933 North Cherry Avenue, Rm. N204, Tucson, AZ 85721-0065, USA}

\author{M. Shahbandeh}
\affiliation{Department of Physics, Florida State University, Tallahassee, FL 32306, USA}

\author{N. B. Suntzeff}
\affiliation{George P. and Cynthia Woods Mitchell Institute for Fundamental Physics \& Astronomy, Texas A\&M University, Department of Physics and Astronomy, 4242 TAMU, College Station, TX 77843}

\author{F. Taddia}
\affiliation{Department of Physics and Astronomy, Aarhus University, 
Ny Munkegade 120, DK-8000 Aarhus C, Denmark }
\affiliation{ The Oskar Klein Centre, Department of Astronomy, Stockholm University, AlbaNova, 106 91 Stockholm, Sweden}

\begin{abstract}


We present the $H$-band wavelength region of  thirty-seven post-maximum light near-infrared (NIR) spectra  of  three normal, nine transitional and four sub-luminous type~Ia supernovae (SNe~Ia), extending from $+$5~d to +20~d relative to  the epoch of $B$-band maximum. 
 We introduce a new observable, the blue-edge velocity, \ved,  of the prominent Fe/Co/Ni-peak $H$-band emission  feature which is quantitatively measured.
 The  \ved\ parameter is found to decrease over sub-type ranging from around  $-$14,000~\kms\  for normal SNe Ia, to $-$10,000~\kms\ for transitional SNe~Ia, down to $-$5,000~\kms\ for the sub-luminous SNe~Ia.
 Furthermore, inspection of the +10$\pm$3\,d  spectra indicates that \ved\ is correlated with 
 the color-stretch parameter, \sBV, and hence with peak luminosity.
These results follow the previous findings that brighter SNe~Ia tend to have \Nifs\ located at higher velocities as  compared to sub-luminous objects. 
As  \ved\ is a model-independent parameter, we propose it can be used in combination with traditional observational diagnostics to provide a new avenue to robustly distinguish between leading SNe~Ia explosion models. 
\end{abstract}


\section{Introduction} 
\label{sect:intro}
Type Ia supernovae (SNe Ia) are the thermonuclear explosions of at least one carbon-oxygen  white dwarf (WD) in a binary system. However the exact nature of their progenitors and critical details of their explosion physics remain open questions. 

The currently favoured progenitor scenarios are the single degenerate scenario (SDS), where a WD accretes material from a non-degenerate companion star such as a H/He or red giant star \citep{Whelan73,Livne90}, and the double degenerate scenario (DDS), consisting of two WDs \citep{Iben84}.  
Within each progenitor scenario there are different explosion mechanisms, and progenitor masses.  However, it is still not clear if all of these scenarios are seen in nature, and if one of them dominates the production of SNe Ia in the Universe.  For a recent review on SNe Ia explosion scenarios see \citet{Livio18}.

\begin{figure*}
\centering
\includegraphics[width=\textwidth]{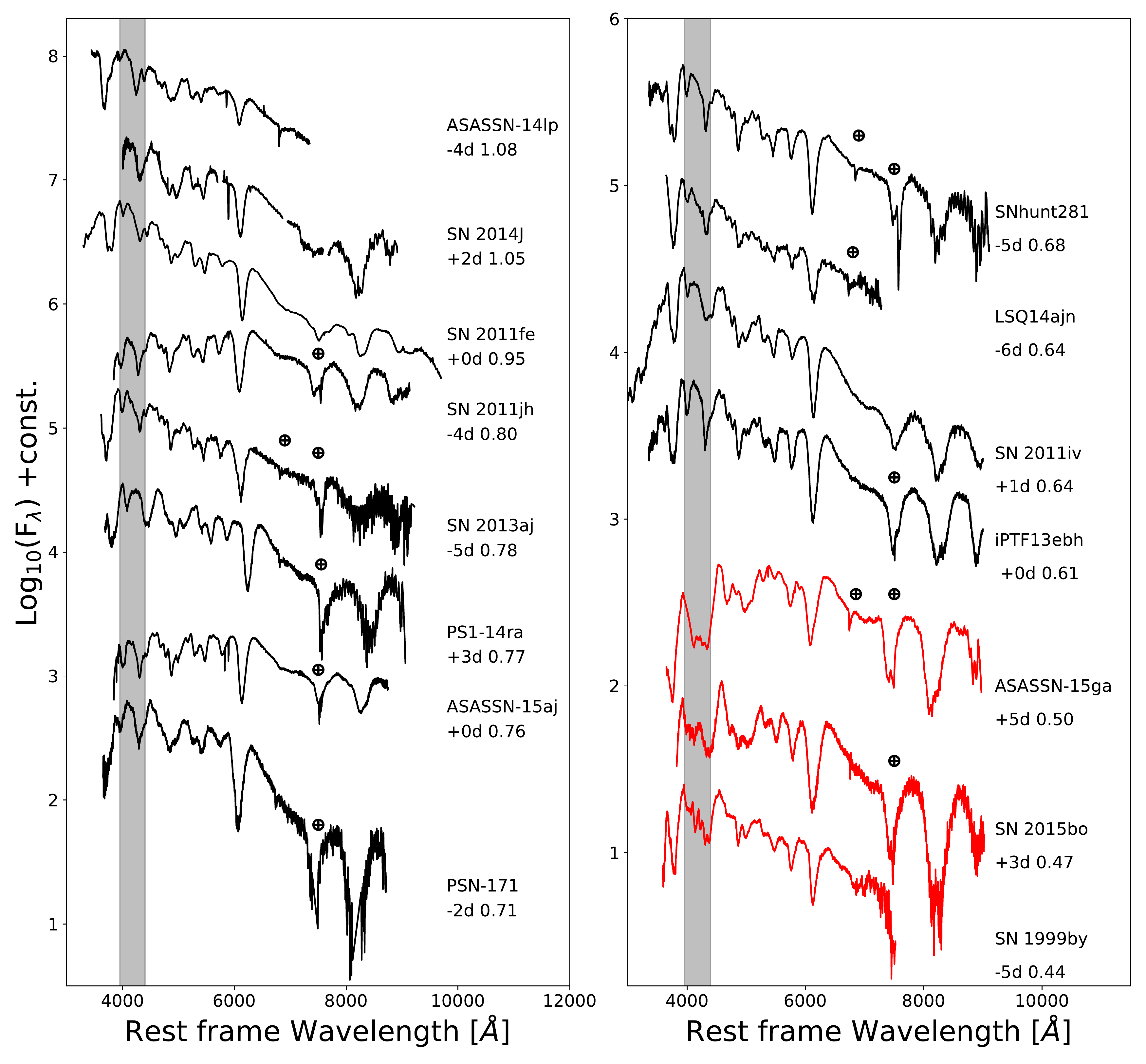}
\caption{Rest-frame visual-wavelength spectra of the SNe used in this work. Their phase relative to maximum and as well as \sBV\ is plotted adjacent to each spectrum.  The objects are plotted in order of \sBV, and those in red  have strong \TiII\ absorption at 4400\AA. The grey vertical region highlights the 
\TiII\ 4400\AA\ feature. All of the spectra are from N. Morrell et al. in prep., except for, ASASSN-14lp \citep{Shappee16}, SN\,2014J \citep{Ashall14}, SN\,2011fe \citep{Pereira13}, SN\,2011iv  \citep{Gall18}, iPTF13ebh  \citep{Hsiao15}, SN~1999by \citep{Garnavich04},  and SN~2014ah \citep{Yaron12}.  The telluric features in the spectra are marked.}
\label{fig:opticalspec}
\end{figure*}

SNe~Ia follow a luminosity-width-relation (LWR), where brighter objects have broader light curves \citep{Phillips93}. Sub-luminous, 1991bg-like  \citep[][]{Filippenko92} SNe~Ia are located at the faint end of the LWR \citep[\eg][]{Ashall16a}, and depending on the parameter combination used, sub-luminous SNe Ia have been proposed to be both part of a continuous distribution from normal SNe Ia \citep[\eg][]{Hoeflich17,Ashall18,Burns18}, and from a distinct population \citep[\eg][]{Stritzinger06,Dhawan17,Blondin17,Scalzo19}.

At NIR wavelengths SNe Ia are nearly standard candles
 and suffer from less systematic effects compared to the optical \citep{Krisciunas04a,WoodV08,Mandel11,Kattner12,Burns14,Dhawan18,Avelino19}.
 As a result current SNe~Ia cosmology programs such as the \textit{Carnegie Supernova Project II} (CSP-II; \citealt{Phillips19}) have turned their attention  towards longer wavelengths. 

NIR spectroscopy offers a promising way to investigate the physics of SNe~Ia,
as it enable us to receive light from different depths in a supernova's atmosphere at the same epoch. This means a single NIR spectrum can simultaneously probe different burning regions in a SN Ia explosion. 
For normal bright SNe~Ia in the NIR, by a few days past maximum light\footnote{Throughout this paper phases (in days) are given with respect to time of rest frame $B$-band maximum.},
there is no well-defined photosphere, and line blanketing dominates the opacity. 
In the region where there are lines, this opacity  provides a quasi-continuum
which is formed at relatively large radii.
Conversely, in areas with few lines there is less opacity, enabling 
much deeper regions  of the ejecta to be visible. 

One wavelength region of interest coincides with the $H$-band where, between maximum light and +10\,d,  a complex iron-peak emission feature emerges due to allowed transitions located above the photosphere  \citep{Kirshner73,Wheeler98,Hoeflich02,Marion09,Hsiao13a}. Previous work has found a correlation between the strength of this feature and the  color-stretch parameter \citep{Hsiao13a}.  
Moreover, with a limited sample, \citet{Hsiao09} found an indication of a correlation between the  velocity of this feature and light curve shape. 
This iron-peak feature consists of a blend of many \FeII/\CoII/\NiII\ allowed emission lines.
The Fe and Co in these epochs are produced through the radioactive decay of \Nifs. 
To first order, the luminosity of a supernova is dependent on the amount of \Nifs\ synthesized in the explosion, where less luminous objects produce smaller amounts of \Nifs\ \citep[\eg][]{Arnett82,Stritzinger06,Mazzali07}.

A primary  objective of the CSP-II was
to obtain a large sample of NIR spectra of SN~Ia \citep{Hsiao19}. In this work, we use a subset of these spectra to examine the $H$-band iron-peak feature in transitional  and sub-luminous SNe~Ia.
Transitional objects are a link between normal-bright  SNe~Ia and the sub-luminous, 1991bg-like population
\citep[see \eg][]{Pastorello07,Hsiao15,Ashall16a,Ashall16b,Gall18}.
Transitional SNe~Ia are characterized by having: i)
 fast declining light curves, with $\DmB>
$1.6\,mag, ii) a secondary $i$-band  NIR maxima which peaks after the time of $B$-band maximum, iii) and no strong \TiII\ absorption at 4400\AA.

Here, we suggest that the highest blue-edge velocity (\ved) of the iron-peak feature represents the outer edge of \Nifs\ in the SNe Ia explosion. In an accompanying  paper \citep{Ashall19}  we  compare  \ved\ to explosion models, and demonstrate it is a measurement of the specific kinetic energy in SNe Ia. We also show that  \ved\ is a quantification of the outer \Nifs\ abundance in the ejecta.
 \ved\ measures the point in velocity space where $X_{\rm{Ni}}$ falls to of order 0.03-0.10.
 Therefore, \ved\ 
  is a direct probe of the sharp transition between the  incomplete and complete Si-burning regions in the ejecta.

Finally, we note that although the light-curve decline-rate parameter-- $\DmB$ 
\footnote{$ \DmB$ is the  difference in magnitude between maximum light and  +15\,d \citep{Phillips93}.} -- has successfully been used to calibrate the luminosity of SNe Ia, it is degenerate when dealing with sub-luminous and transitional objects.  
 This is because, for the least luminous SNe~Ia,
the inflection point in the $B$-band light curve occurs prior to +15\,d \citep{Phillips12}. 
Therefore, throughout the following work we characterize the properties of SNe~Ia light curves with 
the color-stretch \sBV\ parameter. 
\sBV\ is a dimensionless parameter defined as the time difference between $B$-band maximum 
and the reddest point in the $B-V$ color curve divided by 30 days, where typical SNe~Ia have \sBV$\approx$1 \citep{Burns14}.

\begin{figure*}
\centering
\includegraphics[width=\textwidth]{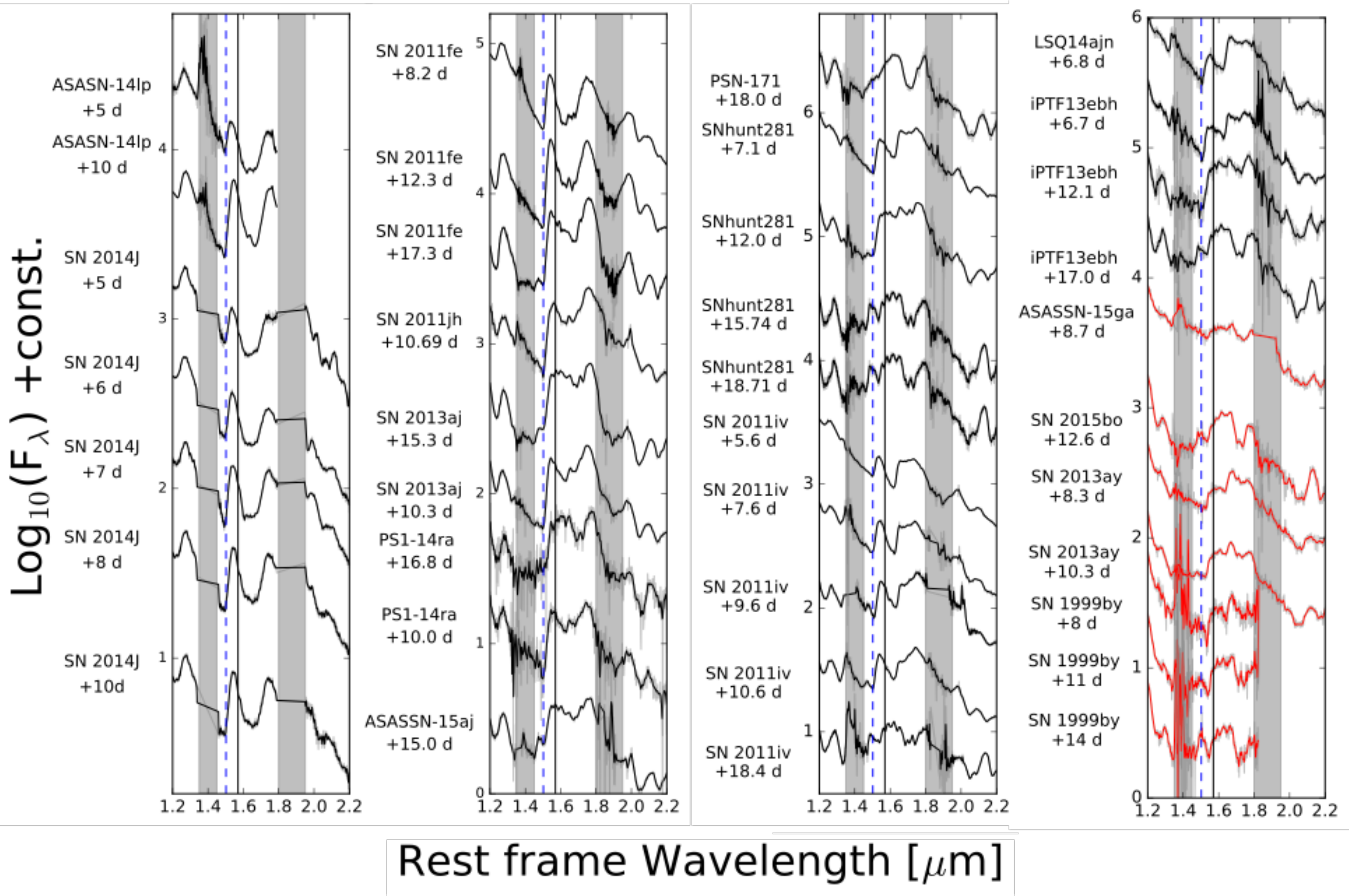}
\caption{The CSP-II SNe~Ia NIR spectra used in this paper, sorted by \sBV, where the objects get fainter from going from the top of the left panel to the bottom of the right panel.  
The name of each supernova and its time relative to $B$-band maximum is adjacent to each spectrum.
The  Gaussian smoothed ($\sigma$=2) spectra are plotted in black and red and the un-smoothed spectra are underneath in grey. The vertical grey regions are the telluric bands in the NIR. The vertical black lines denote the rest wavelength of the 1.57\mic\ feature, and the dashed blue line corresponds to the same feature at -13,000\kms.}
\label{fig:Hsample}
\end{figure*}

\section{Observational Sample}
\label{sect:sample}
\begin{figure*}
\centering
\includegraphics[width=\textwidth]{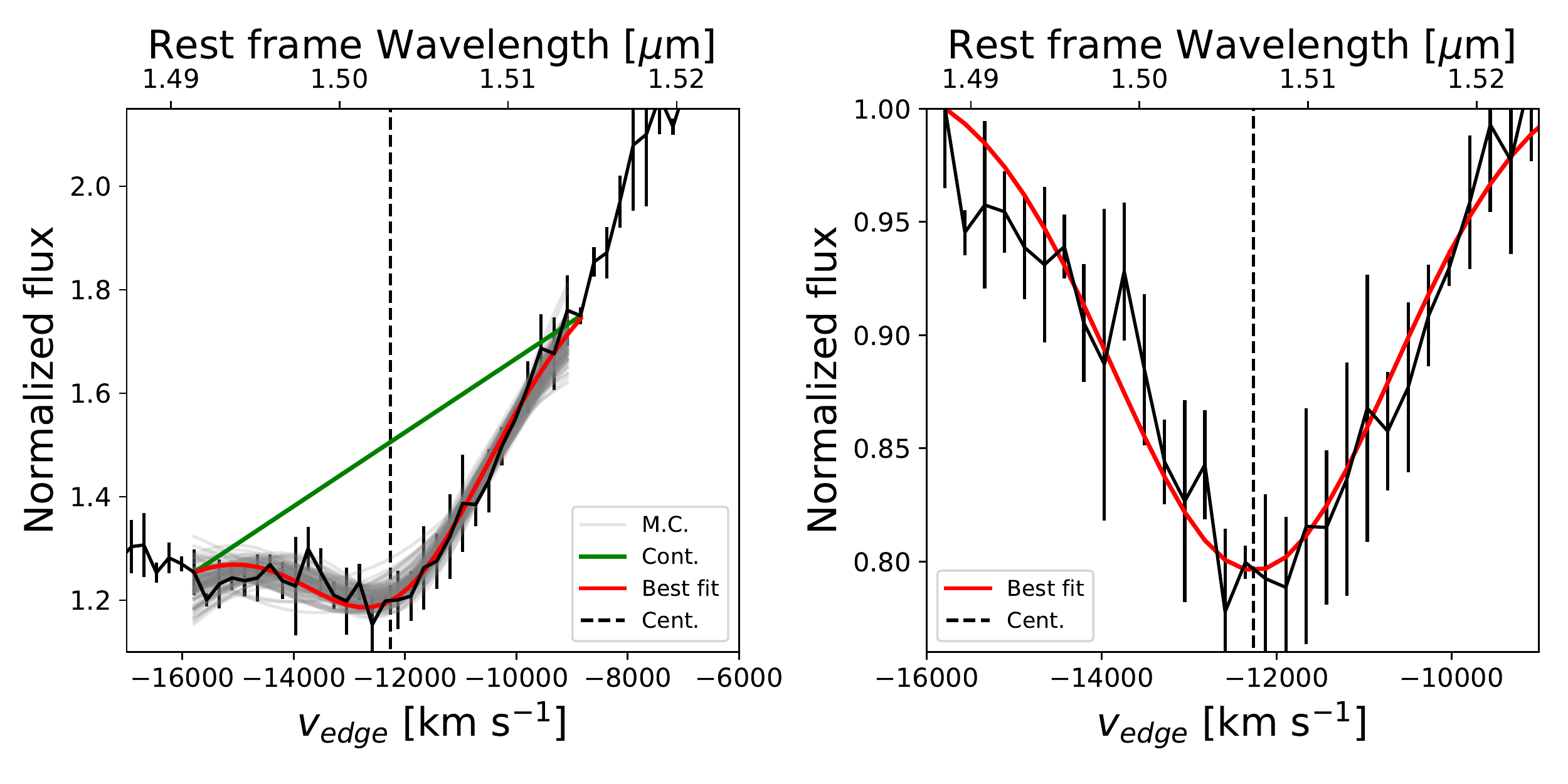}
\caption{An example of how \ved\ is determined using the +7.09\,d spectrum of SNhunt281. 
\textit{Left:} The continuum (green) determined from iteration from the Gaussian fit, the best Gaussian fit (red), the data (black solid), the central value of the Gaussian (black dashed), and the 
100 Monte Carlo realizations (grey). The flux errors of the observations  are provided as error bars. 
\textit{Right:} The best Gaussian fit (red), the normalized data (black solid), and the central value of the Gaussian (black dashed) produced through the fitting process. The continuum from the left panel has been removed. }
\label{fig:examplefit}
\end{figure*}

The selection criteria for the sample were set such that the SNe Ia  were transitional or sub-luminous (i.e., they had \sBV 
$<$0.8), and  have at least one NIR spectrum, between +5 and +20\,d, as these are the phases  when it is predicted that the Fe/Co/Ni emission occurs \citep{Marion09}, and is when the $H$-band break appears in the spectroscopic data. There were 12 SNe Ia that met these criteria from the CSP-II data set. The  sub-luminous SN\,1999by (\sBV=0.44) \citep{Hoeflich02} and the normal SN\,2011fe (\sBV=0.95), SN\,2014J (\sBV=1.05), and ASASSN-14lp (\sBV=1.084)  \citep{Pereira13,Hsiao13a,Marion15} were also used in the analysis.  The basic properties of these SNe are presented in Table \ref{table:NIR}.

Optical spectra near maximum light of our sample (eight of which  are unpublished) are plotted in Fig.~\ref{fig:opticalspec}.
A summary of the optical spectra can be found in Table \ref{table:NIR}. 
The three SNe~Ia plotted in red show the 
 typical characteristic of sub-luminous SNe Ia, a strong \TiII\ absorption feature at \ab4400\AA.
The other SNe Ia have a higher ionization state and are classical transitional objects, except for SN\,2011fe, SN\,2014J and ASASSN-14lp which are normal-bright SNe~Ia. We note that SN\,2013ay was classified from a NIR spectrum \citep{Hsiao13b}, has a light curve shape consistent with a  sub-luminous SNe~Ia and an \sBV=0.46$\pm$0.05, but does not have optical spectra. 
The unpublished CSP-II photometry of all the supernovae was checked,   and the  SNe Ia that were spectroscopically sub-luminous events were found to have small, but not barely visible, 
secondary $i$-band maximum. 
Due to a lack of pre-maximum light curve coverage, for the sub-luminous SNe\,Ia it was not possible to determine 
the $i$-band maximum relative to the $B$-band, except for SN\,2015bo which peaked
in the $i$ band before the $B$-band. 

There are 37 NIR spectra of the 16 SNe~Ia in the sample, 18 of which are unpublished (see Table~\ref{table:NIR} for details).
Most of the spectra were observed with the FIRE spectrograph on the Magellan Baade telescope at Las Campanas Observatory, the other unpublished spectra were observed with SpeX  on the 
NASA Infrared Telescope Facility (IRTF),  Gemini Near-InfraRed Spectrograph (GNIRS) on Gemini-North, and FLAMINGOS-2 on Gemini-South. 
The spectra were reduced and corrected for telluric features via the procedure described by \citet{Hsiao19}. 

The full sample of the NIR spectra is plotted in Fig.~\ref{fig:Hsample}. Each spectrum is corrected to the rest-frame, labeled with the appropriate SN name and rest frame time relative to maximum. 
The figure shows only the $H$-band region of each spectrum.\footnote{The entire NIR wavelength range of these data   will be presented in a future publication.}

\section{Technique}
Here we describe our method to measure \ved.  As demonstrated in Fig.~\ref{fig:examplefit},  \ved\  was measured by fitting the minimum of the region blueward of the iron peak emission feature with a Gaussian profile. The fit is weighted by the observational flux uncertainty. The data were fitted over a \ab0.05$\mu$m range around a central value, illustrated by the red lines in Fig.~\ref{fig:examplefit}.
The continuum is defined as a straight line connecting the end points in this range and removed before the fit.
The Gaussian fit was iterated, using the previous minimum of the 
Gaussian  as the central point in the new fit,  until convergence was met.
This was done to ensure the continumn was properly removed.
Convergence usually required \ab5 iterations. We also fit the data with a Mofat function. Using the Bayesian information criterion, and Akaike information criterion,  it was found that a Gaussian function models the profile better than a  Mofat function.

 An accompanying uncertainty to our best-fit value of \ved\ was determined by producing 100 realizations on a smoothed spectrum 
 with noise added in  at each pixel using a normal distribution with the standard deviation of the Gaussian matching the observed flux error, denoted by the light grey error region in the left panel of Figure \ref{fig:examplefit}. 
 The flux errors of the FIRE spectra come from the standard deviation in flux of the multiple exposures necessary in NIR observations \citep{Hsiao19}.
 This newly constructed spectrum was then measured over the same region and with the same fitting technique as the  observed input spectrum. This was done for each realization to create an array of 100 velocity measurements for the given feature whose standard deviation was taken as the measurement  error of \ved.
 If no error spectrum was available from the observations, an array of values was
 produced through subtracting the smoothed spectrum and the observations. The absolute value of these subtracted values were smoothed to generate a noise trend. 
 This  noise trend was used as a representation of the mean noise on the spectra and sampled 
 back on to the smoothed spectra to produce another realization of the observed spectra. 
 For each of the 100 realizations of the Gaussian fit discussed above, a different amount of noise was sampled from the 
 noise trend back onto the smoothed spectra.

\begin{figure*}
\centering
\includegraphics[scale=0.18]{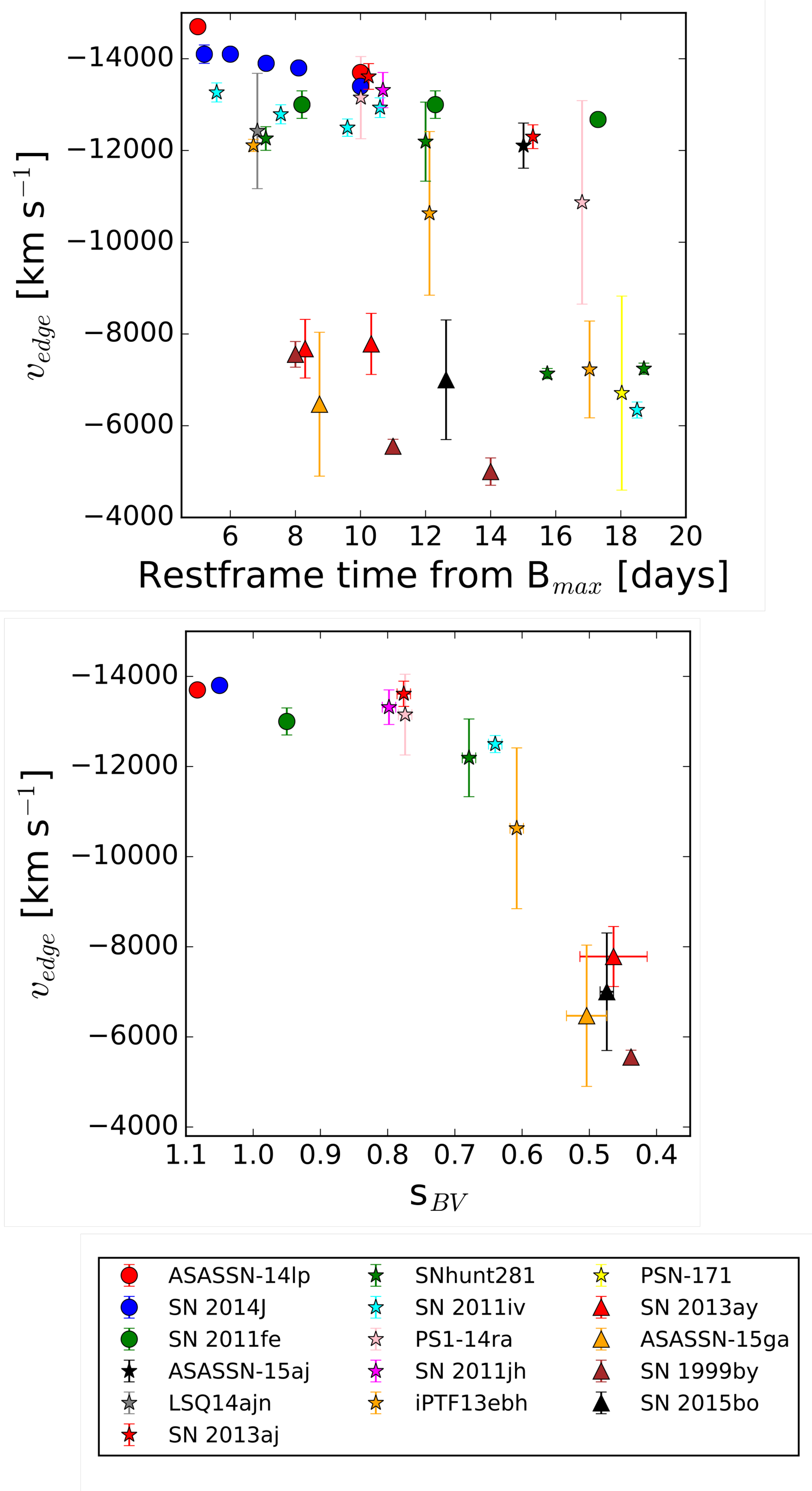}
\caption{\textit{Top:} \ved\  as a function of rest frame time from maximum. \textit{Bottom:} The iron-peak outer velocity at  +10$\pm$3\,d as a function of \sBV. For some objects the error bars are smaller than the marker sizes. Normal SNe Ia are marked by solid
circle symbols, transitional SNe Ia are marked by star circle symbols, and sub-luminous SNe Ia are marked by solid triangle
symbols.}
\label{fig:velall}
\end{figure*}

The iron-peak emission region is a multiplet of many allowed Fe/Co/Ni lines and not from an individual transition. As a result, the different components are sensitive to density and temperature, hence the ionization and excitation state in the centre of the feature can differ.
However,  the Doppler shift of the highest energy (bluest edge) component is not sensitive to density and temperature. This is because it is a result of atomic physics. Therefore we use the outer edge of the feature to measure velocities, as this is a stable measurement. 
In this work we adopt 1.57\mic\ as the rest wavelength of the feature. This 
value was obtained by checking for the strongest lines from non-LTE radiation transport models of normal and sub-luminous SNe Ia  \citep{Hoeflich02,Hoeflich17}. However, if the rest wavelength of this feature were to be altered, it would just cause a systematic shift in the values and not change the trend.

\section{Results}
\label{sect:blItueedge}
Using the method described above, \ved\  was measured for each of the  NIR spectra, and the results are presented in the top panel of  Fig.~\ref{fig:velall}.
Inspection of the measurements  reveals \ved\ decreases over time for the majority of objects.
The normal-bright ASASSN-14lp exhibits the highest velocities, reaching $-14,700\pm$100~\kms\ on +5.2\,d and later $-13,700\pm$100~\kms\ on $+$10.0~d.
Transitional objects  exhibit slightly lower values of \ved, which then decreases rapidly. For example, SN\,2011iv \citep{Gall18,Ashall18} extends from  \ved=$-$13,200$\pm$200~\kms\ at +5.6\,d down to $-6300\pm$200\kms\ by +10\,d.  
The least luminous SNe~Ia have the lowest values of \ved,  clustering around $-6,000$~\kms.
At +8\,d SN\,1999by has a \ved\ of $-7,600\pm$300~\kms\ and by +14\,d it drops to $-5,000\pm$300~\kms.

SNe~Ia with a $v_{edge}$ of
\ab$-$10,000~\kms\ appear to be rare. 
However, SN\,2011iv, SNhunt281 and iPTF13ebh are the only objects which rapidly drops over the time period examined.
For example, iPTF13ebh exhibits a \ved\  of  $-10,600\pm$1,800~\kms\ at +12.12\,d, but by
+17.04\,d  it  drops to $-7,200\pm$1,000~\kms.

In the bottom panel of Fig.~\ref{fig:velall}, we have plotted \ved\ as a function of the color-stretch parameter \sBV. 
The points from the spectra at +10$\pm$3\,d  were plotted because at later times the corresponding layers may become optically thin.
Objects exhibiting  larger \sBV\ values are found to also exhibit higher values of \ved. 
\ved\ was found to range from \ab$-14,000$~\kms\ for normal-bright SNe~Ia, to $-$10,500~\kms\ for transitional SNe~Ia, and down to \ab$-$5,000~\kms\ for sub-luminous SNe~Ia. 
The correlation between \sBV\ and \ved\ implies
\ved\ is also correlated with the peak luminosity of the SNe~Ia, and hence the amount of \Nifs\ produced during the explosion.

From our sample here it is not possible to rule out that there is bimodality in the data, where the sub-luminous SNe~Ia have a larger range. However, photometric properties of SNe~Ia, such as those from CSP-I
show a continuous distribution  from normal to sub-lumious SNe~Ia \citep{Burns18}. This implies that the possible bimodal distribution seen in this work is produced by a small sample size, and the fact that transitional SNe Ia are rare.
A Pearson correlation test of the points in the bottom panel of Fig.~\ref{fig:velall} produced a correlation coefficient of -0.86.

\section{Discussion}
\label{sect:conclusion}
A measure of \ved\  provides a constraint on of the  outer \Nifs\ distribution in the ejecta.
This is due to the fact that the  Fe/Co/Ni emission region is located at  large radii in the SN atmosphere is an isolated multiplet and is not contaminated by lines from other elements.
Furthermore, the highest energy Doppler shift of this region  will correspond to the edge of the highest velocity emission, which is  a measurement of the  outer \Nifs. 
\ved\ is therefore an valuable tool to analyze the explosion physics of  SNe\,Ia. 
For example measuring  \ved\ for  a large sample of SNe will probe the mixing of \Nifs\ in the ejecta.

\ved\ is  a model-independent measurement, and a diagnostic of the location between the complete and incomplete Si-burning regions.
It provides an  indication of the effectiveness of the burning in the ejecta.
 Those objects that are brighter, and produce more \Nifs, have experienced more effective burning. The exact details of this will be explored in an accompanying paper  with respect to explosion models \citep{Ashall19}. However, the results here are consistent with those found through nebular phase spectral modeling by \citet{Mazzali98} who showed that less luminous objects have \Nifs\ located at lower velocities. A similar result was also found by \citet{Boty17} who demonstrate using nebular phase modeling that a larger \Nifs\ mass (and therefore SN luminosity) produces broader Fe lines in the ejecta. 

Our results indicate that \Nifs\ is located at lower velocities for less luminous SNe~Ia, and  may argue
against very low mass explosions (less than \ab0.9\Msun) for the sub-luminous SNe~Ia. 
This is because, for the same \Nifs\  mass, these very low mass explosions tend to have the
\Nifs\ located at higher velocities compared to $M_{Ch}$ explosions \citep[\eg][]{Sim10,Blondin18}, which may be in contradiction with the values of \ved\ seen here.  
 Finally, the fact that we see a relatively smooth distribution in \ved\ as a  function of light curve shape implies that at least some SNe~Ia at the faint end of the luminosity width relationship may have a similar explosion mechanism to normal SNe Ia.


\vspace{0.5cm}
\acknowledgements
We  thank many colleagues for fruitful  discussions, as well as   Josh Simon, Povilas Palunas, and Mansi Kasliwal  for providing telescope time.
The work presented here has been supported 
in part by NSF awards AST-1008343 \& AST-1613426 (PI: M.M. Phillips), AST-1613472 (PI: E.Y Hsiao), AST-1715133 (PI: P. Hoeflich), AST-
16113455 (PI: N. B. Suntzeff) and in part by a Sapere Aude Level II grant (PI: M.D. Stritzinger) from  the Danish National Research Foundation (DFF).
M.D. Stritzinger and S. Holmbo are generously supported by a research grant (13261) from the VILLUM FONDEN.  M.D. Stritzinger and E. Baron acknowledge support from  the Aarhus University Research Fund (AUFF) for  a Sabbatical research grant. E. Baron acknowledges partial support from NASA Grant NNX16AB25G. Research by D.J. Sand is supported by NSF grants AST-1821967, 1821987, 1813708 and 1813466.
We thank the Mitchell Institute for Fundamental Physics and Astronomy for partial support.
Based on observations obtained at the Gemini Observatory under program  GS-2015A-Q-5 (PI: D. Sand).  Gemini is operated by the Association of Universities for Research in Astronomy, Inc., under a cooperative agreement with the NSF on
behalf of the Gemini partnership: the NSF (United States), the National
Research Council (Canada), CONICYT (Chile), Ministerio de Ciencia, Tecnolog\'ia e Innovaci\'on Productiva (Argentina), and Minist\'erio da Ci\^encia, Tecnologia e Inova\c{c}\~ao (Brazil).
D. J. Sand is a visiting Astronomer at the Infrared Telescope Facility, which is operated by the University of Hawaii under contract NNH14CK55B with the National Aeronautics and Space Administration.
Based in part on observations made with the Nordic Optical Telescope (P49-017, P50-015, and P51-006; PI M.D. Stritzinger), operated by the Nordic Optical Telescope Scientific Association at the Observatorio del Roque de los Muchachos, La Palma, Spain, of the Instituto de Astrofisica de Canarias.

Facilities: Magellan:Baade (FIRE near-infrared echellette), Magellan:Clay (Magellan Inamori Kyocera Echelle), du Pont (Boller \& Chivens spectrograph, Gemini:North (GNIRS near-infrared spectrograph), VLT (ISAAC), NTT (EFOSC), IRTF (SpeX near-infrared spectrograph), Hale (DBSP), Hiltner (TIFKAM), NOT (ALFOSC), UH:2.2m (SNIFS), FLWO:1.5m (FAST), HST (STIS), La Silla-QUEST, CRTS, iPTF, ASAS- SN, PS1)

Facilities: Python, IRAF, IDL

\startlongtable
\begin{deluxetable*}{ccccccccc}
\tablewidth{\textwidth}
\tablecaption{The properties of the SNe Ia and a log of the NIR and optical spectral observations. The $\DmB$ values were obtained from a direct spline fit to the data, and the values of  \sBV were obtained from the best fits from SNooPy. \label{table:NIR}}
\tablehead{
\colhead{SN}&
\colhead{z}&
\colhead{\sBV}&
\colhead{$\DmB$}&
\colhead{$T_{Bmax}$\footnote{Time $B$-band maximum, calculated from the CSP-II light curves.}}&
\colhead{$T_{spec}$\footnote{Time of spectral observation.}}&
\colhead{Phase\footnote{ \label{FN:P} Phase of spectra in rest frame relative to $B$-band maximum.}}&
\colhead{\ved}&
\colhead{Instrument/Telescope}\\
\colhead{}&
\colhead{}&
\colhead{}&
\colhead{mag}&
\colhead{JD$-$2,450,000}&
\colhead{JD$-$2,450,000}&
\colhead{days}&
\colhead{\kms}&
\colhead{}}
\startdata
  \multicolumn{9}{c}{\bf NIR}\\ 
 \hline
    ASASSN-14lp &0.005&1.08$\pm$0.05&0.85$\pm$0.07&57015.3&57020.3&+5.0&$-$14700$\pm$100&F2/Gemini-S\\
    &&&&&57025.3&+10.0&$-$13700$\pm$100&F2/Gemini-S\\
   SN 2014J&0.0001&1.05$\pm$0.086&1.05$\pm$0.05& 56689.75&56694.95&+5.2&$-$14,100$\pm$200&Mt Abu\footnote{\label{FN:M15} \citet{Marion15}}\\
   &&&&&56695.78&+6.0&$-$14,100$\pm$100&Mt Abu\textsuperscript{\ref{FN:M15}}\\
   &&&&&56696.93&+7.1&$-$13,900$\pm$100&Mt Abu\textsuperscript{\ref{FN:M15}}\\
   &&&&&56697.92&+8.1&$-$13,800$\pm$100&Mt Abu\textsuperscript{\ref{FN:M15}}\\
   &&&&&56699.84&+10.0&$-$13,400$\pm$100&Mt Abu\textsuperscript{\ref{FN:M15}}\\
 SN\,2011fe&0.001&0.95$\pm$0.01&1.21$\pm$0.05&55813.93&55822.13&+8.2&$-$13,800$\pm$600&GNIRS/Gemini-N\\
&&&&&55826.22&+12.3&$-$13,500$\pm$100&GNIRS/Gemini-N \\
&&&&&55831.21&+17.3&$-$13,300$\pm$100&GNIRS/Gemini-N\\
 SN\,2011jh&0.008&0.80$\pm$0.01&1.46$\pm$0.01&55931.06&55941.84&+10.69*&$-$13,300$\pm$400&FIRE/Baade\\
 SN\,2013aj&0.009&0.78$\pm$0.01&1.47$\pm$0.01&56361.37&56371.72&+10.25*&$-$13,600$\pm$300&FIRE/Baade\\
&&&&&56376.81&+15.30*&$-$12,300$\pm$300&FIRE/Baade\\
 PS1-14ra&0.028&0.77$\pm$0.01&$\cdots$&56724.54&56734.84&+10.01*&$-$13,000$\pm$900&FIRE/Baade\\
&&&&&56741.82&+16.81*&$-$10,900$\pm$2200&FIRE/Baade\\
 ASASSN-15aj&0.011&0.76$\pm$0.01&1.44$\pm$0.02&57035.46&57050.64&+15.01*&$-$12,100$\pm$500&FIRE/Baade\\
 PSN-171 \footnote{\label{FN:PSN171} J13471211-2422171}&0.020&0.71$\pm$0.01&1.54$\pm$0.02&57070.43&57088.82&+18.03*&$-$6,700$\pm$2000&FIRE/Baade\\\
 SNhunt281\footnote{ \label{FN:SNhunt281}SN\,2015bp}&0.004&0.68$\pm$0.01&1.56$\pm$0.03&57112.67&57119.79&+7.09*&$-$12,200$\pm$300&FIRE/Baade\\
&&&&&57124.72&+12.00*&$-$12,200$\pm$900&FIRE/Baade\\
&&&&&57128.41&+15.74*&$-$7,100$\pm$100&GNIRS/Gemini-N\\
&&&&&57131.28&+18.71*&$-$7,200$\pm$100&Spex/IRTF\\
 LSQ14ajn\footnote{\label{FN:LSQ14ajn} SN\,2014ah}&0.021&0.64$\pm$0.01&1.74$\pm$0.02&56734.73&56741.70&+6.83*&$-$12,400$\pm$1200&FIRE/Baade\\
 SN\,2011iv&0.006&0.64$\pm$0.01&1.74$\pm$0.01&55906.08&55911.70&+5.58&$-$13,200$\pm$200&FIRE/Baade\\
&&&&&55913.68&+7.55&$-$12,800$\pm$200&FIRE/Baade\\
&&&&&55915.7&+9.60&$-$12,500$\pm$200& SOFI/NTT\\
 &&&&&55916.75&+10.60&$-$12,900$\pm$200&FIRE/Baade\\
 &&&&&55924.6&+18.50&$-$6,300$\pm$200& ISAAC/VLT\\
iPTF13ebh&0.013&0.61$\pm$0.01&1.76$\pm$0.02&56623.29&56630.0&+6.71&$-$12,100$\pm$100&GNIRS/Gemini-N\\
  &&&&&56635.58&+12.12&$-$10,600$\pm$1800 &FIRE/Baade\\
 &&&&&56640.56&+17.04&$-$7,200$\pm$1000&FIRE/Baade\\
ASASSN-15ga&0.007&0.50$\pm$0.03&2.13$\pm$0.04&57115.88&57124.68&+8.74*&$-$6,500$\pm$1500&FIRE/Baade\\
 SN\,2015bo&0.016&0.47$\pm$0.01&1.85$\pm$0.02&57076.02&57088.86&+12.63*&$-$7,000$\pm$1300&FIRE/Baade\\
 SN 2013ay&0.016&0.46$\pm$0.05&$\cdots$&56375.41&56383.84&+8.30*&$-$7,700$\pm$600&FIRE/Baade\\
&&&&&56385.91&+10.33*&$-$7,800$\pm$700&FIRE/Baade\\
  SN\,1999by&0.002&0.44$\pm$0.01&1.90$\pm$0.05&51308&51316&+8&$-$7,600$\pm$300&TIFKAM/Hiltner\footnote{\citet{Hoeflich02}}\\
&&&&&51319&+11&$-$5,500$\pm$200&TIFKAM/Hiltner\\
&&&&&51322&+14&$-$5,000$\pm$300&TIFKAM/Hiltner\\
 \hline
  \multicolumn{9}{c}{\bf Optical}\\ 
   \hline
   ASASSN-14lp &0.005&1.08$\pm$0.05&0.85$\pm$0.07&57015.3&57011&-4&$\cdots$&FLWO-1.5m / FAST\footnote{\citet{Shappee16}}\\
   SN 2014J&0.0001&1.05$\pm$0.086&1.05$\pm$0.05& 56689.75&56692&+2&$\cdots$&Frodospec/LT\footnote{\citet{Ashall14}}\\
   SN 2011fe&0.001&0.95$\pm$0.01&1.21$\pm$0.05&55813.93  &55814&+0&$\cdots$&SNIFS/UH 2.2m\footnote{\citet{Pereira13}}\\
   SN 2011jh&0.008&0.80$\pm$0.01&1.46$\pm$0.01&55931.06&55927&$-$4&$\cdots$&B\&C/Du Pont\footnote{ \label{FN:Mor} Morrell et al. in prep.}\\
   SN 2013aj&0.009&0.78$\pm$0.01&1.47$\pm$0.01&56361.37&56356&$-$5&$\cdots$&EFOSC/NTT\footnote{\citet{Smartt15}}\\
   PS1-14ra&0.028&0.77$\pm$0.01&$\cdots$&56724.54&56728&+3&$\cdots$&ALFOSC/NOT\textsuperscript{\ref{FN:Mor}}\\
   ASASSN-15aj&0.011&0.76$\pm$0.01&1.44$\pm$0.02&57035.46&57035&+0&$\cdots$&MIKE Clay \textsuperscript{\ref{FN:Mor}}\\
   PSN-171\textsuperscript{\ref{FN:PSN171}}&0.020&0.71$\pm$0.01&1.54$\pm$0.02&57070.43&57068&$-$2&$\cdots$&ALFOSC/NOT\textsuperscript{\ref{FN:Mor}}\\
  SNhunt281\textsuperscript{\ref{FN:SNhunt281}}&0.004&0.68$\pm$0.01&1.56$\pm$0.03&57112.67&57108&$-$5&$\cdots$&ALFOSC/NOT\textsuperscript{\ref{FN:Mor}}\\
   LSQ14ajn\textsuperscript{\ref{FN:LSQ14ajn}}&0.021&0.64$\pm$0.01&1.74$\pm$0.02&56734.73 &56729&$-$6&$\cdots$&ALFOSC/NOT\textsuperscript{\ref{FN:Mor}}\\
   SN 2011iv&0.006&0.64$\pm$0.01&1.74$\pm$0.01&55906.08&55907&+1&$\cdots$&STIS/HST\footnote{\citet{Gall18}}\\
   iPTF13ebh&0.013&0.61$\pm$0.01&1.76$\pm$0.02&56623.29&56623&+0&$\cdots$&DBSP/Palomar200\footnote{\citet{Hsiao15}}\\
   ASASSN-15ga&0.007&0.50$\pm$0.03&2.13$\pm$0.04&57115.88&57121&+5&$\cdots$&ALFOSC/NOT\textsuperscript{\ref{FN:Mor}}\\
   SN 2015bo&0.016&0.47$\pm$0.01&1.85$\pm$0.02&57076.02 &57079&+3&$\cdots$&B\&C/Du Pont\textsuperscript{\ref{FN:Mor}}\\
   SN 1999by&0.002&0.44$\pm$0.01&1.90$\pm$0.05&51308&51303&$-$5&$\cdots$&FAST/FLWO\footnote{\citet{Garnavich04}}
\enddata
\end{deluxetable*}

\acknowledgments

\end{document}